\documentclass[nofootinbib,prd]{revtex4}%
\usepackage{amsmath}
\usepackage{amsfonts}
\usepackage{amssymb}
\usepackage{graphicx}%
\begin{document}
\title{Effects of an Electric Charge on Casimir Wormholes: Changing the Throat Size}
\author{Remo Garattini}
\email{remo.garattini@unibg.it}
\affiliation{Universit\`{a} degli Studi di Bergamo, Dipartimento di Ingegneria e Scienze
Applicate,Viale Marconi 5, 24044 Dalmine (Bergamo) Italy and }
\affiliation{I.N.F.N. - sezione di Milano, Milan, Italy.}

\begin{abstract}
In this paper we continue the investigation about the connection between
Casimir energy and the traversability of a wormhole. In addition to the
negative energy density obtained by a Casimir device, we include the effect of
an electromagnetic field generated by an electric charge. This combination
defines an electrovacuum source which has an extra parameter related to the
size of the throat. Even if the electromagnetic energy density is positive the
Null Energy Condition is still violated. The main reason is that the
electromagnetic field satisfies the property $\rho=-p_{r}$. As a consequence
the Traversable Wormhole throat can be changed as a function of the electric
charge. This means that the throat is no more Planckian and the traversability
is little less in principle but little more in practice.

\end{abstract}
\maketitle

\section{Introduction}

Casimir wormholes are Traversable Wormholes (TW) obtained by solving the
semiclassical Einstein's Field Equations (EFE)%
\begin{equation}
G_{\mu\nu}=\kappa\left\langle T_{\mu\nu}\right\rangle ^{\text{Ren}}%
\qquad\kappa=\frac{8\pi G}{c^{4}}%
\end{equation}
with a source of the form%
\begin{equation}
\rho_{C}\left(  d\right)  =-\frac{\hbar c\pi^{2}}{720d^{4}},\label{rhoC}%
\end{equation}%
\begin{equation}
P\left(  d\right)  =\frac{F\left(  d\right)  }{S}=-3\frac{\hbar c\pi^{2}%
}{720d^{4}},\label{P(a)}%
\end{equation}
representing the energy density and the pressure, respectively. $d$ is the
plates separation\cite{CW}. $\left\langle T_{\mu\nu}\right\rangle
^{\text{Ren}}$ describes the renormalized stress-energy tensor of some matter
fields which, in this specific case, is obtained by the Zero Point Energy
(ZPE) contribution of the electromagnetic field. The two key ingredients
useful to form a Casimir wormhole are in the relationship $P\left(  d\right)
/\rho_{C}\left(  d\right)  =3$. This particular number is the cornerstone of a
Casimir wormhole. In addition, for a Casimir wormhole, the plates separation
$d$ has been promoted to be the radial coordinate $r$ dealt with a variable.
For this reason the pressure $P\left(  d\right)  $ will be interpreted as a
radial pressure $p_{r}\left(  r\right)  $ To build a Casimir wormhole we need
to introduce the following spacetime metric%
\begin{equation}
ds^{2}=-e^{2\Phi(r)}\,dt^{2}+\frac{dr^{2}}{1-b(r)/r}+r^{2}d\Omega
^{2}\,,\label{metric}%
\end{equation}
where $d\Omega^{2}=d\theta^{2}+\sin^{2}\theta d\phi^{2}$ is the line element
of the unit sphere. $\Phi(r)$ and $b(r)$ are two arbitrary functions of the
radial coordinate $r\in\left[  r_{0},+\infty\right)  $, denoted as the
redshift function and the shape function, respectively \cite{MT,MTY,Visser}.
With the help of the metric $\left(  \ref{metric}\right)  $ the EFE, written
in an orthonormal frame, are%
\begin{equation}
\frac{b^{\prime}\left(  r\right)  }{r^{2}}=\kappa\rho\left(  r\right)
,\label{rhoEFE}%
\end{equation}%
\begin{equation}
\frac{2}{r}\left(  1-\frac{b\left(  r\right)  }{r}\right)  \Phi^{\prime
}(r)-\frac{b\left(  r\right)  }{r^{3}}=\kappa p_{r}\left(  r\right)
\label{pr0}%
\end{equation}
and%
\begin{align}
&  \Bigg\{\left(  1-\frac{b\left(  r\right)  }{r}\right)  \left[  \Phi
^{\prime\prime}(r)+\Phi^{\prime}(r)\left(  \Phi^{\prime}(r)+\frac{1}%
{r}\right)  \right]  \nonumber\\
&  -\frac{b^{\prime}\left(  r\right)  r-b\left(  r\right)  }{2r^{2}}\left(
\Phi^{\prime}(r)+\frac{1}{r}\right)  \Bigg\}=\kappa p_{t}(r).\label{pt0}%
\end{align}
$\rho\left(  r\right)  $ is the energy density\footnote{However, if
$\rho\left(  r\right)  $ represents the mass density, then we have to replace
$\rho\left(  r\right)  $ with $\rho\left(  r\right)  c^{2}.$}, $p_{r}\left(
r\right)  $ is the radial pressure, and $p_{t}\left(  r\right)  $ is the
lateral pressure. The line element $\left(  \ref{metric}\right)  $ represents
a spherically symmetric and static wormhole and $r_{0}$ is the location of the
throat. We can complete the EFE with the expression of the conservation of the
stress-energy tensor which can be written in the same orthonormal reference
frame%
\begin{equation}
p_{r}^{\prime}\left(  r\right)  =\frac{2}{r}\left(  p_{t}\left(  r\right)
-p_{r}\left(  r\right)  \right)  -\left(  \rho\left(  r\right)  +p_{r}\left(
r\right)  \right)  \Phi^{\prime}(r).\label{Tmn}%
\end{equation}
If we assume that an Equation of State (EoS) $p_{r}\left(  r\right)
=\omega\rho_{C}\left(  r\right)  $ is imposed, then we find the following
solution to the semiclassical EFE%
\begin{align}
\Phi(r) &  =\frac{1}{2}\left(  {\omega-1}\right)  {\ln}\left(  {\frac{r\omega
}{\left(  \omega r+r_{0}\right)  }}\right)  \\
b(r) &  =\left(  1-\frac{1}{\omega}\right)  r_{0}+\frac{r_{0}^{2}}{\omega r},
\end{align}
where we have used the energy density $\rho_{C}\left(  r\right)  $ of
Eq.$\left(  \ref{rhoC}\right)  $ and the radial pressure $p_{r}\left(
r\right)  $ described by Eq.$\left(  \ref{P(a)}\right)  $ as a source with $d$
replaced by $r$. A fundamental property of a wormhole is that a flaring out
condition of the throat, given by $(b-b^{\prime}r)/b^{2}>0$, must be satisfied
as well as the request that $1-b(r)/r>0$. Furthermore, at the throat
$b(r_{0})=r_{0}$ and the condition $b^{\prime}(r_{0})<1$ is imposed to have
wormhole solutions. It is also fundamental that there are no horizons present,
which are identified as the surfaces with $e^{2\Phi(r)}\rightarrow0$, so that
$\Phi(r)$ must be finite everywhere. The procedure used to obtain a Casimir
wormhole can be extended to include also the electromagnetic field as
additional source. The key point is in the following observation: the
algebraic structure of stress-energy tensors for electromagnetic fields is
determined by\cite{Dymnikova,DG}%
\begin{equation}
T_{0}^{0}=T_{1}^{1}%
\end{equation}
that it means%
\begin{equation}
\rho=-p_{r}.\label{prho}%
\end{equation}
As a warm up exercise we can consider a pure spherically symmetric
electromagnetic field without the contribution of the Casimir energy, to see
if it is possible to build a TW even if the energy density is positive. In an
orthonormal frame, the SET we are going to consider is the following%
\begin{equation}
T_{\mu\nu}^{EM}=\frac{Q^{2}}{2\left(  4\pi\right)  ^{2}\varepsilon_{0}r^{4}%
}diag(1,-1,1,1)
\end{equation}
which is conserved and traceless. From the energy density $\rho$ of the SET,
it is possible to obtain the following shape function%
\begin{equation}
b\left(  r\right)  =r_{0}+\frac{r_{2}^{2}}{r_{0}}-\frac{r_{2}^{2}}{r};\qquad
r_{2}^{2}=\frac{GQ^{2}}{4\pi c^{4}\varepsilon_{0}},\label{b(r)E}%
\end{equation}
which has the correct properties. Indeed%
\begin{equation}
b^{\prime}\left(  r_{0}\right)  =\frac{r_{2}^{2}}{r_{0}^{2}}<1\qquad
\Longrightarrow\qquad r_{2}<r_{0}\label{fo}%
\end{equation}
and%
\begin{align}
1-\frac{b(r)}{r} &  =\frac{\left(  r-r_{0}\right)  \left(  rr_{0}-r_{2}%
^{2}\right)  }{r_{0}r}>0\qquad if\qquad r>r_{0};r>\frac{r_{2}^{2}}{r_{0}}\\
\qquad\frac{b-b^{\prime}r}{b^{2}} &  =\frac{r_{0}r^{2}\left(  r_{0}^{2}%
+r_{2}^{2}\right)  }{\left(  r_{0}^{2}r+r_{2}^{2}r-r_{2}^{2}r_{0}\right)
^{2}}>0.
\end{align}
Eq.$\left(  \ref{pr0}\right)  $ together with the shape function $\left(
\ref{b(r)E}\right)  $ allow the computation of the redshift function%
\begin{equation}
\Phi(r)=\frac{1}{2}\ln\left(  \frac{\left(  r-r_{0}\right)  \left(
rr_{0}-r_{2}^{2}\right)  }{r^{2}}\right)  +C{,}%
\end{equation}
where $C$ is an integration constant\footnote{Actually, because of the
property $\left(  \ref{prho}\right)  $, from Eq.$\left(  \ref{pr0}\right)  $,
one can write%
\begin{equation}
\frac{2}{r}\left(  1-\frac{b\left(  r\right)  }{r}\right)  \Phi^{\prime
}(r)-\frac{b\left(  r\right)  }{r^{3}}+\frac{b^{\prime}\left(  r\right)
}{r^{2}}=0
\end{equation}
which has the following solution%
\begin{equation}
\Phi(r)=\frac{1}{2}\ln\left(  1-\frac{b\left(  r\right)  }{r}\right)  +C,
\end{equation}
which is a signature of a Black Hole.}. As expected, we cannot adopt the
strategy of Ref.\cite{CW} because the $\ln\left(  r-r_{0}\right)  $ never
disappears for every choice of $r_{0}$, even if property $\left(
\ref{prho}\right)  $ is satisfied. One could insist by imposing an
inhomogeneous EoS of the form%
\begin{equation}
\omega\left(  r\right)  =-\frac{b\left(  r\right)  }{rb^{\prime}\left(
r\right)  }=\frac{r_{0}r_{2}^{2}-\left(  r_{0}^{2}+r_{2}^{2}\right)  r}%
{r_{2}^{2}r_{0}}\label{o1}%
\end{equation}
allowing to fix $\Phi(r)=0$. Nevertheless, Eq.$\left(  \ref{o1}\right)  $ must
be compatible with%
\begin{equation}
\omega\left(  r\right)  =\frac{p_{r}\left(  r\right)  }{\rho\left(  r\right)
}=-1.\label{o2}%
\end{equation}
This leads to%
\begin{equation}
\omega\left(  r_{0}\right)  =-1
\end{equation}
which is incompatible with the flare out condition $\left(  \ref{fo}\right)
$, because one gets%
\begin{equation}
r_{0}=r_{2}.
\end{equation}
This means that a pure electromagnetic field cannot support a TW even if
$\rho+p_{r}=0$. It is necessary to have $\rho+p_{r}<0$. For this reason, we
are led to consider the superposition of the Casimir source with the
electromagnetic field. Such a combination potentially could produce a
different result thanks to the property $\left(  \ref{prho}\right)  $ which
defines an electrovacuum source. Note that such an electrovacuum source has
been investigated earlier in Ref.\cite{STP} even in the context of G.U.P.
distortions. Note also that the idea of including an electric charge or an
electromagnetic field in a TW configuration is not new. Indeed, the first
proposal is due to Kim and Lee\cite{KimLee} who considered a combination
between the Morris-Thorne wormhole and the Reissner-Nordstr\"{o}m spacetime.
Balakin et al. discussed a nonminimal Einstein-Maxwell model \cite{BLZ}.
Kuhfittig\cite{Kuhfittig} considered a modification of the Kim and Lee charged
wormhole to have compatibility with the quantum inequality of Ford and
Roman\cite{FR}. The purpose of this paper is to repeat the same procedure
which has led to the original Casimir wormhole spacetime to see if it is
possible to obtain new EFE solutions with an additional electric field. The
rest of the paper is structured as follows, in section \ref{p2} we continue
the investigation to determine if the Casimir energy density with an
additional electric charge can be considered as a source for a traversable
wormhole, in section \ref{p3} we examine the features of the Casimir wormhole
with the additional electric charge, in section \ref{p4} we consider the
Casimir energy and the additional electric charge with the plates separation
regarded as a parameter instead of a variable. We summarize and conclude in
section \ref{p5}.

\section{The Casimir Traversable Wormhole with an Additional Electric Charge}

\label{p2}In this section we assume that our exotic matter will be represented
by the Casimir energy density $\left(  \ref{rhoC}\right)  $. Following
Ref.\cite{CW}, we promote the constant plates separation $d$ to a radial
coordinate $r$. In addition to the Casimir source we include the contribution
of an electric field generated by a point charge. Since it is the NEC that
must be violated, the following inequality $\rho\left(  r\right)
+p_{r}\left(  r\right)  <0$ must hold. We want to draw the reader's attention
that, thanks to the property $\left(  \ref{prho}\right)  $, we can write%
\begin{equation}
\rho\left(  r\right)  +p_{r}\left(  r\right)  =\rho_{C}\left(  r\right)
+p_{r,C}\left(  r\right)  +\rho_{E}\left(  r\right)  +p_{r,E}\left(  r\right)
=-\frac{4\hbar c\pi^{2}}{720r^{4}}<0,
\end{equation}
where%
\begin{equation}
\rho_{C}\left(  r\right)  =-\frac{\hbar c\pi^{2}}{720r^{4}};\qquad
p_{r,C}\left(  r\right)  =-\frac{3\hbar c\pi^{2}}{720r^{4}};\qquad\rho
_{E}\left(  r\right)  =\frac{Q^{2}}{2\left(  4\pi\right)  ^{2}\varepsilon
_{0}r^{4}};\qquad p_{r,E}\left(  r\right)  =-\frac{Q^{2}}{2\left(
4\pi\right)  ^{2}\varepsilon_{0}r^{4}}.
\end{equation}
In this context, the total energy density is represented by%
\begin{equation}
\rho\left(  r\right)  =\rho_{C}\left(  r\right)  +\rho_{E}\left(  r\right)
=-\frac{\hbar c\pi^{2}}{720r^{4}}+\frac{Q^{2}}{2\left(  4\pi\right)
^{2}\varepsilon_{0}r^{4}}=-\frac{r_{1}^{2}}{\kappa r^{4}}+\frac{r_{2}^{2}%
}{\kappa r^{4}},\label{rho(r)}%
\end{equation}
where%
\begin{align}
r_{1}^{2} &  =\frac{\pi^{3}l_{p}^{2}}{90},\label{r1}\\
r_{2}^{2} &  =\frac{GQ^{2}}{4\pi c^{4}\varepsilon_{0}}.\label{r2}%
\end{align}
Thus%
\begin{equation}%
\begin{array}
[c]{c}%
\rho\left(  r\right)  <0\\
\rho\left(  r\right)  =0\\
\rho\left(  r\right)  >0
\end{array}
\qquad when\qquad%
\begin{array}
[c]{c}%
r_{1}>r_{2}\\
r_{1}=r_{2}\\
r_{1}<r_{2}%
\end{array}
.
\end{equation}
The shape function $b\left(  r\right)  $ can be obtained plugging $\rho\left(
r\right)  $ of Eq.$\left(  \ref{rho(r)}\right)  $ into Eq.$\left(
\ref{rhoEFE}\right)  $, whose solution leads to%
\begin{equation}
b\left(  r\right)  =r_{0}+\left(  r_{2}^{2}-r_{1}^{2}\right)  \int_{r_{0}}%
^{r}\frac{dr^{\prime}}{r^{\prime2}}=r_{0}-\frac{r_{1}^{2}-r_{2}^{2}}{r_{0}%
}+\frac{r_{1}^{2}-r_{2}^{2}}{r},\label{b(r)0}%
\end{equation}
where the throat condition $b(r_{0})=r_{0}$ has been imposed. The redshift
function can be obtained by solving Eq.$\left(  \ref{pr0}\right)  $ with the
help of the shape function $\left(  \ref{b(r)0}\right)  $. One finds%
\begin{equation}
\frac{2}{r}\left(  1-\frac{r_{0}}{r}+\frac{r_{1}^{2}-r_{2}^{2}}{r_{0}r}%
-\frac{r_{1}^{2}-r_{2}^{2}}{r^{2}}\right)  \Phi^{\prime}\!\left(  r\right)
-\frac{r_{0}}{r^{3}}+\frac{r_{1}^{2}-r_{2}^{2}}{r_{0}r^{3}}-\left(
1-\omega\right)  \frac{r_{1}^{2}-r_{2}^{2}}{r^{4}}=0,
\end{equation}
where we have used an EoS of the form $p_{r}\left(  r\right)  =\omega
\rho\left(  r\right)  $. The solution can be written as%
\begin{gather}
\Phi\left(  r\right)  =\left(  \frac{r_{0}^{2}-\omega\left(  r_{1}^{2}%
-r_{2}^{2}\right)  }{2\left(  r_{0}^{2}+r_{1}^{2}-r_{2}^{2}\right)  }\right)
\ln\!\left(  r-r_{0}\right)  \nonumber\\
-\left(  \frac{\omega r_{0}^{2}-r_{1}^{2}+r_{2}^{2}}{2\left(  r_{0}^{2}%
+r_{1}^{2}-r_{2}^{2}\right)  }\right)  \ln\!\left(  rr_{0}+r_{1}^{2}-r_{2}%
^{2}\right)  +\frac{\omega-1}{2}\ln\!\left(  r\right)  +C.\label{Phi(r)}%
\end{gather}
If $Q=0$, one recovers the familiar form of the Casimir wormhole redshift
function. It is possible to eliminate the horizon if we constrain $\omega$ to
be%
\begin{equation}
\omega=\omega_{0}=\frac{r_{0}^{2}}{r_{1}^{2}-r_{2}^{2}}\label{o}%
\end{equation}
and the redshift function $\left(  \ref{Phi(r)}\right)  $ becomes%
\begin{equation}
\Phi\left(  r\right)  =-\left(  \frac{\omega-1}{2}\right)  \ln\!\left(
\frac{\omega rr_{0}+r_{0}^{2}}{\omega r}\right)  +C.
\end{equation}
By assuming that $\Phi\left(  r\right)  \rightarrow0$ for $r\rightarrow\infty
$, then we find%
\begin{equation}
\Phi\left(  r\right)  =\frac{\omega-1}{2}\ln\!\left(  \frac{\omega r}{\omega
r+r_{0}}\right)  ,
\end{equation}
which is formally the same result of Ref.\cite{CW}. The shape function
$\left(  \ref{b(r)0}\right)  $ can be rearranged to obtain the familiar form
of the Casimir wormhole\cite{CW}
\begin{equation}
b\!\left(  r\right)  =r_{0}-\frac{r_{1}^{2}-r_{2}^{2}}{r_{0}}+\frac{r_{1}%
^{2}-r_{2}^{2}}{r}=r_{0}\left(  1-\frac{1}{\omega}\right)  +\frac{r_{0}^{2}%
}{\omega r}.\label{b(r)00}%
\end{equation}
On the other hand, since the ratio in Eq.$\left(  \ref{o}\right)  $ is not
constrained, we can use the ratio $p_{r}\left(  r\right)  /\rho\left(
r\right)  $ to extract information about the size of the throat. It is
immediate to see that for%
\begin{equation}
\omega=\frac{p_{r}\left(  r\right)  }{\rho\left(  r\right)  }=\frac{3r_{1}%
^{2}+r_{2}^{2}}{r_{1}^{2}-r_{2}^{2}};\qquad r_{1}\neq r_{2},\label{or1r2}%
\end{equation}
the EoS is satisfied and is independent on $r$. Plugging the value of $\omega$
in Eq.$\left(  \ref{or1r2}\right)  $ into Eq.$\left(  \ref{o}\right)  $, one
finds
\begin{equation}
r_{0}=\sqrt{3r_{1}^{2}+r_{2}^{2}}.\label{r0}%
\end{equation}
Note that $r_{2}$ can be variable, while $r_{1}$ is not. It is convenient to
take $r_{1}$ as a reference scale. Thus, if we introduce a dimensionless
variable%
\begin{equation}
x=\frac{r_{2}}{r_{1}}=\sqrt{\frac{90GQ^{2}}{4\pi c^{4}\varepsilon_{0}\pi
^{3}l_{p}^{2}}}=\sqrt{\frac{90}{\pi^{3}}n^{2}\frac{e^{2}}{4\pi\varepsilon
_{0}\hbar c}}=\frac{3n}{\pi}\sqrt{\frac{10}{\pi}\alpha},\label{fine}%
\end{equation}
one finds%
\begin{equation}
\omega=\frac{3+x^{2}}{1-x^{2}};\qquad r_{0}=r_{1}\sqrt{3+x^{2}};\qquad x\neq1,
\end{equation}
where $Q=ne$, $e$ is the electron charge, $\alpha$ is the fine structure
constant and $n$ is the total number of the electron charges. Note that, for%
\begin{equation}
Q\rightarrow0,\qquad r_{2}\rightarrow0\qquad\Longrightarrow x\rightarrow0,
\end{equation}
one recovers the shape function of Ref.\cite{CW} with $\omega=3$. On the other
hand, when%
\begin{equation}
Q\rightarrow\infty,\qquad r_{2}\rightarrow\infty\qquad\Longrightarrow
x\rightarrow\infty\label{Limit}%
\end{equation}
and $\omega\rightarrow-1$. In conclusion, we can write%
\begin{align}
\omega &  \in\left(  3,+\infty\right)  ;\qquad x\in\left(  0,1^{-}\right)  \\
\omega &  \in\left(  -\infty,-1\right)  ;\qquad x\in\left(  1^{+}%
,+\infty\right)  .
\end{align}
A comment on the limit $\left(  \ref{Limit}\right)  $ is in order. Indeed,
when $\omega\rightarrow-1$, the shape function assumes the form%
\begin{equation}
b\left(  r\right)  =2r_{0}-\frac{r_{0}^{2}}{r}%
\end{equation}
and the redshift function is%
\begin{equation}
\Phi\left(  r\right)  =\ln\!\left(  \frac{r-r_{0}}{r}\right)
\end{equation}
representing no more a TW, but a black hole. However, $\omega\rightarrow-1$ is
a limiting value which never will be reached. This means that $Q$ can be
arbitrarily large but finite. Note that in this range $\omega<-1$. This means
that the pure electromagnetic field, in this context, acts as a phantom energy
source. For completeness, we report the expression of the transverse pressure
which, in terms of $\omega$, is identical to $p_{t}\left(  r\right)  $ of
Ref.\cite{CW}. One finds%
\begin{equation}
p_{t}\left(  r\right)  =\frac{r_{0}^{2}}{\kappa\omega{r}^{4}}\left[
\omega+\frac{r_{0}\left(  1-{\omega}^{2}\right)  }{4\omega\,\left(
\omega\,r+r_{0}\right)  }\right]  =\omega_{t}\left(  r\right)  \left(
\frac{r_{0}^{2}}{\kappa\omega\,{r}^{4}}\right)  =\omega_{t}\left(  r\right)
\rho(r),
\end{equation}
where we have introduced an inhomogeneous EoS on the transverse pressure with%
\begin{equation}
\omega_{t}\left(  r\right)  =-\left(  \omega+\frac{r_{0}\left(  1-{\omega}%
^{2}\right)  }{4\omega\,\left(  \omega\,r+r_{0}\right)  }\right)  ,
\end{equation}
and the final form of the SET is%
\begin{equation}
T_{\mu\nu}=\frac{r_{0}^{2}}{\kappa\omega\,{r}^{4}}\left[  diag\left(
-1,-\omega,\omega_{t}\left(  r\right)  ,\omega_{t}\left(  r\right)  \right)
\right]  .
\end{equation}
The conservation of the SET is satisfied but a comparison with the SET source
shows that%
\begin{equation}
T_{\mu\nu}=T_{\mu\nu}^{Source}-\frac{1}{\kappa r^{4}}\left[  diag\left(
0,0,\omega_{t}\left(  r\right)  -\left(  r_{1}^{2}+r_{2}^{2}\right)
,\omega_{t}\left(  r\right)  -\left(  r_{1}^{2}+r_{2}^{2}\right)  \right)
\right]  ,
\end{equation}
namely a discrepancy on the transverse pressure with respect to the
SET\ source is present. We recall that the SET\ source is defined by%
\begin{equation}
T_{\mu\nu}^{Source}=T_{\mu\nu}^{Casimir}+T_{\mu\nu}^{EM}=\frac{1}{\kappa
r^{4}}\left[  diag\left(  -r_{1}^{2}+r_{2}^{2},-\left(  3r_{1}^{2}+r_{2}%
^{2}\right)  ,r_{1}^{2}+r_{2}^{2},r_{1}^{2}+r_{2}^{2}\right)  \right]  .
\end{equation}
It is important to observe that there exists another interesting value for
$\omega$, namely when $\omega=1$. For this special choice, one finds that the
line element reduces to the Ellis-Bronnikov (EB)
wormhole\cite{ellisGL,Bronnikov}, namely%
\begin{equation}
ds^{2}=-\,dt^{2}+\frac{dr^{2}}{1-\frac{r_{0}^{2}}{r^{2}}}+r^{2}\,d\Omega^{2}%
\end{equation}
whose associated SET is%
\begin{equation}
T_{\mu\nu}^{EB}=\frac{r_{1}^{2}-r_{2}^{2}}{\kappa{r}^{4}}\left[  diag\left(
-1,-1,1,1\right)  \right]  .
\end{equation}
Nevertheless $\omega=1$ is incompatible with the relationship $\left(
\ref{or1r2}\right)  $ and therefore this option will be discarded.

\subsection{Special case $r_{1}^{2}=r_{2}^{2}$}

\label{p2a}In the special case%
\begin{equation}
r_{1}^{2}=r_{2}^{2}=r_{e}^{2},
\end{equation}
we find that the energy density vanishes. Therefore the wormhole shape
function is%
\begin{equation}
b\left(  r\right)  =r_{0}.
\end{equation}
On the other hand the redshift function appears to be non trivial. Indeed,
from the EFE $\left(  \ref{pr0}\right)  $, we find%
\begin{equation}
\frac{2}{r}\left(  1-\frac{r_{0}}{r}\right)  \Phi^{\prime}\!\left(  r\right)
-\frac{r_{0}}{r^{3}}+\frac{4r_{e}^{2}}{r^{4}}=0
\end{equation}
which can be rearranged to give%
\begin{equation}
\Phi^{\prime}\!\left(  r\right)  =\frac{r_{0}r-4r_{e}^{2}}{2\left(
r-r_{0}\right)  r^{2}}.
\end{equation}
The solution is%
\begin{equation}
\Phi\left(  r\right)  =-\frac{\ln\!\left(  r\right)  }{2}+\frac{2\ln\!\left(
r\right)  r_{e}^{2}}{r_{0}^{2}}-\frac{2r_{e}^{2}}{r_{0}r}+\frac{\ln\!\left(
r-r_{0}\right)  }{2}\left(  1-\frac{4r_{e}^{2}}{r_{0}^{2}}\right)  +C.
\end{equation}
It is immediate to see that for%
\begin{equation}
r_{0}=2r_{e}%
\end{equation}
the horizon disappears and%
\begin{equation}
\Phi\left(  r\right)  =-\frac{r_{0}}{2r},
\end{equation}
where we have assumed that $\Phi\left(  r\right)  \rightarrow0$ for
$r\rightarrow\infty$. Therefore in this special case we still have a TW with
the following line element%
\begin{equation}
ds^{2}=-\exp\left(  -\frac{r_{0}}{r}\right)  \,dt^{2}+\frac{dr^{2}}{1-r_{0}%
/r}+r^{2}d\Omega^{2}\,, \label{scase}%
\end{equation}
which is traversable in principle but not in practice because the throat is
Planckian. To complete this special case, we compute the transverse pressure
and we find%
\begin{equation}
\Bigg\{\frac{r_{0}^{2}}{r^{4}}-\frac{r_{0}^{3}}{4r^{5}}\Bigg\}=\kappa
p_{t}(r).
\end{equation}
Note that for this special case, we cannot impose an EoS of the form
$p_{r}\left(  r\right)  =\omega\rho\left(  r\right)  $ because $\rho\left(
r\right)  $ is vanishing. In the next section, we are going to explore some of
the features of the TW obtained by the Casimir source and the electromagnetic field.

\section{Properties of the Casimir wormhole with an additional electromagnetic
field}

\label{p3}In section \ref{p2}, we have introduced the shape function $\left(
\ref{b(r)0}\right)  $ or $\left(  \ref{b(r)00}\right)  $ obtained by the
Casimir energy plus the electromagnetic field. Here we want to discuss some of
its properties. The first quantity we are going to analyze is the proper
radial distance, defined by%
\begin{equation}
l\left(  r\right)  =\pm\int_{r_{0}}^{r}\frac{dr^{\prime}}{\sqrt{1-\frac
{b\left(  r^{\prime}\right)  }{r^{\prime}}}}. \label{l(r)}%
\end{equation}
In this specific case, plugging Eq.$\left(  \ref{b(r)00}\right)  $ into
Eq.$\left(  \ref{l(r)}\right)  $, one gets%
\begin{gather}
l\left(  r\right)  =\pm\int_{r_{0}}^{r}\frac{dr^{\prime}}{\sqrt{1-\frac{r_{0}%
}{r^{\prime}}\left(  1-\frac{1}{\omega}\right)  -\frac{r_{0}^{2}}{\omega
r^{\prime2}}}}=\pm\left(  \frac{\sqrt{r-r_{0}}\,\sqrt{\omega r+r_{0}}\,}%
{\sqrt{\omega}}\right. \nonumber\\
\left.  +r_{0}\frac{\omega-1}{2\omega}\ln\!\left(  \frac{r_{0}+\left(
2r-r_{0}\right)  \omega+2\sqrt{r-r_{0}}\,\sqrt{\omega r+r_{0}}\,\sqrt{\omega}%
}{\!\left(  \omega+1\right)  r_{0}}\right)  \right)  . \label{l(r)1}%
\end{gather}
We find%
\begin{equation}
l\left(  r\right)  \simeq%
\begin{array}
[c]{ll}%
\pm\left(  r+\frac{r_{0}\left(  \omega-1\right)  }{2\omega}\left(
\ln\!\left(  \frac{4\omega r}{r_{0}\left(  \omega+1\right)  }\right)
\!-1\right)  +O\left(  \frac{1}{r}\right)  \right)  & r\rightarrow\infty\\
\pm2\sqrt{\frac{r_{0}\omega}{\omega+1}\left(  r-r_{0}\right)  }+O\left(
\left(  r-r_{0}\right)  ^{\frac{3}{2}}\right)  & r\rightarrow r_{0}%
\end{array}
, \label{l(r)1a}%
\end{equation}
where the\textquotedblleft$\pm$\textquotedblright\ depends on the wormhole
side we are. The proper radial distance is an essential tool to estimate the
possible time trip in going from one station located in the lower universe,
say at $l=-l_{1}$, and ending up in the upper universe station, say at
$l=l_{2}$. Following Ref.\cite{MT}, we shall locate $l_{1}$ and $l_{2}$ at a
value of the radius such that $l_{1}\simeq l_{2}\simeq10^{4}r_{0}$ that it
means $1-b\left(  r\right)  /r\simeq1$. Assume that the traveller has a radial
velocity $v\left(  r\right)  $, as measured by a static observer positioned at
$r$. One may relate the proper distance travelled $dl$, radius travelled $dr$,
coordinate time lapse $dt$, and proper time lapse as measured by the observer
$d\tau$, by the following relationships%
\begin{equation}
v=e^{-\Phi\left(  r\right)  }\frac{dl}{dt}=e^{-\Phi\left(  r\right)  }\left(
1-\frac{b\left(  r\right)  }{r}\right)  ^{-\frac{1}{2}}\frac{dr}{dt}
\label{vt}%
\end{equation}
and%
\begin{equation}
v\gamma=\frac{dl}{d\tau}=\mp\left(  1-\frac{b\left(  r\right)  }{r}\right)
^{-\frac{1}{2}}\frac{dr}{d\tau};\qquad\gamma=\left(  1-\frac{v^{2}\left(
r\right)  }{c^{2}}\right)  ^{-\frac{1}{2}} \label{vtau}%
\end{equation}
respectively. If the traveler journeys with constant speed $v$, then the total
time is given by%
\begin{equation}
\Delta t=\int_{r_{0}}^{r}\frac{e^{-\Phi\left(  r^{\prime}\right)  }dr^{\prime
}}{v\sqrt{1-\frac{b\left(  r^{\prime}\right)  }{r^{\prime}}}}=\omega
^{1-\frac{\omega}{2}}\int_{r_{0}}^{r}\frac{r^{^{\prime}\frac{3}{2}%
-\frac{\omega}{2}}\left(  \omega r^{\prime}+r_{0}\right)  ^{-1+\frac{\omega
}{2}}}{v\sqrt{r^{\prime}-r_{0}}}dr^{\prime} \label{Deltat}%
\end{equation}
while the proper total time is%
\begin{gather}
\Delta\tau=\int_{r_{0}}^{r}\frac{dr^{\prime}}{v\sqrt{1-\frac{b\left(
r^{\prime}\right)  }{r^{\prime}}}}=\frac{\sqrt{r-r_{0}}\,\sqrt{\omega r+r_{0}%
}\,}{v\sqrt{\omega}}\nonumber\\
+r_{0}\frac{\omega-1}{2\omega}\ln\!\left(  \frac{r_{0}+\left(  2r-r_{0}%
\right)  \omega+2\sqrt{r-r_{0}}\,\sqrt{\omega r+r_{0}}\,\sqrt{\omega}%
}{\!\left(  \omega+1\right)  r_{0}}\right)  .
\end{gather}
To evaluate $\Delta t$ we can proceed with the following approximations. Close
to the throat, one finds%
\begin{align}
\Delta t  &  \simeq\omega^{1-\frac{\omega}{2}}\int_{r_{0}}^{r}\frac
{r^{^{\prime}\frac{3}{2}-\frac{\omega}{2}}\left(  \omega r^{\prime}%
+r_{0}\right)  ^{-1+\frac{\omega}{2}}}{v\sqrt{r^{\prime}-r_{0}}}dr^{\prime
}\nonumber\\
&  \simeq\frac{2\sqrt{r_{0}}}{v}\left(  \frac{\omega}{\omega+1}\right)
^{1-\frac{\omega}{2}}\sqrt{r-r_{0}}%
\end{align}
and, in this range, $\Delta t\simeq\Delta\tau$ except for the value
$\omega=-1$, where the TW turns into a Black Hole. On the other hand, when
$r\rightarrow\infty$, one finds%
\begin{equation}
\Delta t\simeq\int_{r_{0}}^{r}\frac{r^{^{\prime}\frac{1}{2}}}{v\sqrt
{r^{\prime}-r_{0}}}dr^{\prime}\simeq\frac{r_{0}}{v}\left(  1-\frac{1}{\omega
}\right)  \ln\!\left(  \frac{r}{r_{0}}\right)  +\frac{r}{v}%
\end{equation}
and even with this approximation, the leading term is the same of $\Delta\tau
$. On the same ground, we can compute the embedded surface, which is defined
by%
\begin{equation}
z\left(  r\right)  =\pm\int_{r_{0}}^{r}\frac{dr^{\prime}}{\sqrt{\frac
{r^{\prime}}{b\left(  r^{\prime}\right)  }-1}}%
\end{equation}
and, in the present case, we find%
\begin{gather}
z\left(  r\right)  =\pm\int_{r_{0}}^{r}\frac{\sqrt{r_{0}}\,\sqrt{\left(
\omega-1\right)  r^{\prime}+r_{0}}}{\sqrt{\omega r^{\prime}+r_{0}}%
\,\sqrt{r^{\prime}-r_{0}}}dr^{\prime}\nonumber\\
\pm\frac{2r_{0}}{\omega^{2}}\left(  F\!\left(  \frac{\sqrt{r-r_{0}}%
\,\sqrt{\omega}}{\sqrt{\omega r+r_{0}}},\frac{1}{\omega}\right)  +\Pi\!\left(
\frac{\sqrt{r-r_{0}}\,\sqrt{\omega}}{\sqrt{\omega r+r_{0}}},1,\frac{1}{\omega
}\right)  \left(  \omega^{2}-1\right)  \right)  ,
\end{gather}
where $F\left(  \varphi,k\right)  $ is the elliptic integral of the first kind
and $\Pi\left(  \varphi,n,k\right)  $ is the elliptic integral of the third
kind. Close to the throat, one can write%
\begin{equation}
z\left(  r\right)  \simeq\pm2\sqrt{r_{0}}\sqrt{\frac{\omega}{1+\omega}}%
\sqrt{r-r_{0}}.
\end{equation}
It is interesting to note the singularity appearing when $\omega=-1$, showing
the presence of a Black Hole. To further investigate the properties of the
shape function $\left(  \ref{b(r)00}\right)  $, we consider the computation of
the total gravitational energy for a wormhole\cite{NZCP}, defined as%
\begin{gather}
E_{G}\left(  r\right)  =\int_{r_{0}}^{r}\left[  1-\sqrt{\frac{1}{1-b\left(
r^{\prime}\right)  /r^{\prime}}}\right]  \rho\left(  r^{\prime}\right)
dr^{\prime}r^{\prime2}\nonumber\\
+\frac{r_{0}}{2G}=\left(  M-M_{\pm}^{P}\right)  c^{2},
\end{gather}
where $M$ is the total mass $M$ and $M^{P}$ is the proper mass, respectively.
Even in this case, the \textquotedblleft$\pm$\textquotedblright\ depends one
the wormhole side we are. In particular%
\begin{gather}
M=\int_{r_{0}}^{r}\frac{4\pi}{c^{2}}\rho\left(  r^{\prime}\right)  r^{\prime
2}dr^{\prime}+\frac{r_{0}}{2Gc^{2}}=\frac{c^{2}}{2G}\left(  r_{0}\left(
1-\frac{1}{\omega}\right)  +\frac{r_{0}^{2}}{\omega r}-r_{0}\right)
+\frac{r_{0}c^{2}}{2G}\nonumber\\
=\frac{c^{2}}{2G}\left(  -\frac{r_{0}}{\omega}+\frac{r_{0}^{2}}{\omega
r}\right)  +\frac{r_{0}}{2Gc^{2}}\underset{r\rightarrow\infty}{\simeq}%
-\frac{r_{0}c^{2}}{2G\omega}+\frac{r_{0}c^{2}}{2G}=\frac{r_{0}c^{2}}%
{2G}\left(  1-\frac{1}{\omega}\right)
\end{gather}
and%
\begin{align}
M_{\pm}^{P}  &  =\pm\frac{4\pi}{c^{2}}\int_{r_{0}}^{r}\frac{\rho\left(
r^{\prime}\right)  r^{\prime2}}{\sqrt{1-b\left(  r^{\prime}\right)
/r^{\prime}}}dr^{\prime}=\pm\frac{c^{2}}{2G}\int_{r_{0}}^{r}\frac{b^{\prime
}\left(  r^{\prime}\right)  }{\sqrt{1-b\left(  r^{\prime}\right)  /r^{\prime}%
}}dr^{\prime}\nonumber\\
\mp &  \frac{r_{0}c^{2}}{4G\sqrt{\omega}}\left(  \pi-2\arctan\!\left(
\frac{r\omega-r+2r_{0}}{2\sqrt{r\omega+r_{0}}\,\sqrt{r-r_{0}}}\right)  \right)
\nonumber\\
&  \underset{r\rightarrow\infty}{\simeq}\mp\frac{r_{0}c^{2}}{4G\sqrt{\omega}%
}\left(  \pi-2\arctan\!\left(  \frac{\omega-1}{2\sqrt{\omega}}\right)
\right)  .
\end{align}
Thus, at infinity one finds%
\begin{equation}
E_{G}\left(  r\right)  \underset{r\rightarrow\infty}{\simeq}\frac{r_{0}c^{2}%
}{2G}\left[  \left(  1-\frac{1}{\omega}\right)  \mp\frac{1}{2\sqrt{\omega}%
}\left(  \pi-2\arctan\!\left(  \frac{\omega-1}{2\sqrt{\omega}}\right)
\right)  \right]  .
\end{equation}
An important traversability condition is that the acceleration felt by the
traveller should not exceed Earth's gravity $g_{\oplus}\simeq980$ $cm/s^{2}$.
In an orthonormal basis of the traveller's proper reference frame, we can find%
\begin{equation}
\left\vert \mathbf{a}\right\vert =\left\vert \sqrt{1-\frac{b\left(  r\right)
}{r}}e^{-\Phi\left(  r\right)  }\left(  \gamma e^{\Phi\left(  r\right)
}\right)  ^{\prime}\right\vert \leq\frac{g_{\oplus}}{c^{2}}. \label{acc}%
\end{equation}
If we assume a constant speed and $\gamma\simeq1$, then we can write%
\begin{equation}
\left\vert \mathbf{a}\right\vert =\left\vert \sqrt{1-\frac{r_{0}}{r}\left(
1-\frac{1}{\omega}\right)  -\frac{r_{0}^{2}}{\omega r^{2}}}\frac{\left(
\omega-1\right)  r_{0}}{2r\left(  \omega r+r_{0}\right)  }\right\vert
\leq\frac{g_{\oplus}}{c^{2}}.
\end{equation}
We can see that in proximity of the throat, the traveller has a vanishing
acceleration. Always following Ref.\cite{MT}, we can estimate the tidal forces
by imposing an upper bound represented by $g_{\oplus}$. The radial tidal
constraint
\begin{gather}
\left\vert \left(  1-\frac{b\left(  r\right)  }{r}\right)  \left[
\Phi^{\prime\prime}\left(  r\right)  +\left(  \Phi^{\prime}\left(  r\right)
\right)  ^{2}-\frac{b^{\prime}\left(  r\right)  r-b\left(  r\right)
}{2r\left(  r-b\left(  r\right)  \right)  }\Phi^{\prime}\left(  r\right)
\right]  \right\vert \times\nonumber\\
c^{2}\left\vert \eta^{\hat{1}^{\prime}}\right\vert \leq g_{\oplus},
\label{RTC}%
\end{gather}
constrains the redshift function, and the lateral tidal constraint%
\begin{gather}
\left\vert \frac{\gamma^{2}c^{2}}{2r^{2}}\left[  \frac{v^{2}\left(  r\right)
}{c^{2}}\left(  b^{\prime}\left(  r\right)  -\frac{b\left(  r\right)  }%
{r}\right)  +2r\left(  r-b\left(  r\right)  \right)  \Phi^{\prime}\left(
r\right)  \right]  \right\vert \times\nonumber\\
\left\vert \eta^{\hat{2}^{\prime}}\right\vert \leq g_{\oplus}, \label{LTC}%
\end{gather}
constrains the velocity with which observers traverse the wormhole.
$\eta^{\hat{1}^{\prime}}$ and $\eta^{\hat{2}^{\prime}}$ represent the size of
the traveller. In Ref.\cite{MT}, they are fixed approximately equal, at the
symbolic value of $2$ $m$. Close to the throat, the radial tidal constraint
$\left(  \ref{RTC}\right)  $ becomes%
\begin{gather}
\left\vert \left[  \frac{b\left(  r\right)  -b^{\prime}\left(  r\right)
r}{2r^{2}}\Phi^{\prime}\left(  r\right)  \right]  \right\vert \nonumber\\
=\frac{\left(  \left(  \omega-1\right)  r+2r_{0}\right)  r_{0}^{2}\left(
\omega-1\right)  }{4\omega\left(  \omega r+r_{0}\right)  r^{4}}\underset
{r\rightarrow r_{0}}{=}\frac{1}{6r_{0}^{2}}\leq\frac{g_{\oplus}}%
{c^{2}\left\vert \eta^{\hat{1}^{\prime}}\right\vert }\nonumber\\
\qquad\Longrightarrow\qquad10^{8}m\lesssim r_{0}. \label{RTCt}%
\end{gather}
For the lateral tidal constraint, we find%
\begin{gather}
\frac{v^{2}r_{0}}{2r^{4}}\left\vert \frac{r\left(  \omega-1\right)  +2r_{0}%
}{\omega}\right\vert \left\vert \eta^{\hat{2}^{\prime}}\right\vert \lesssim
g_{\oplus}\qquad\Longrightarrow\qquad v\lesssim r_{0}\sqrt{\left\vert
\frac{\omega}{\omega+1}\right\vert g_{\oplus}}\nonumber\\
\qquad\Longrightarrow\qquad v\lesssim3.13r_{0}\sqrt{\left\vert \frac{\omega
}{\omega+1}\right\vert }\text{ }m/s. \label{LTCt}%
\end{gather}
If the observer has a vanishing $v$, then the tidal forces are null. We can
use these last estimates to complete the evaluation of the crossing time which
approximately is%
\begin{equation}
\Delta t\simeq\frac{\Delta l}{3.13r_{0}\sqrt{\left\vert \frac{\omega}%
{\omega+1}\right\vert }}\left(  \frac{\omega}{\omega+1}\right)  ^{-\frac
{\omega}{2}}\simeq6.4\times10^{3}\left(  \left\vert \frac{\omega}{\omega
+1}\right\vert \right)  ^{-\frac{1}{2}}\left(  \frac{\omega}{\omega+1}\right)
^{-\frac{\omega}{2}}s,
\end{equation}
which is in agreement with the estimates found in Ref.\cite{MT}, even for
$\omega\rightarrow\pm\infty$. The last property we are going to discuss is the
\textquotedblleft total amount\textquotedblright\ of ANEC violating matter in
the spacetime\cite{VKD} which is described by%
\begin{equation}
I_{V}=\int[\rho(r)+p_{r}(r)]dV
\end{equation}
and for the line element $\left(  \ref{metric}\right)  $, one can write%
\begin{equation}
I_{V}=\frac{1}{\kappa}\int\left(  r-b\left(  r\right)  \right)  \left[
\ln\left(  \frac{e^{2\Phi(r)}}{1-\frac{b\left(  r\right)  }{r}}\right)
\right]  ^{\prime}dr, \label{IV}%
\end{equation}
where the measure $dV$ has been changed into $r^{2}dr$. For the metric
$\left(  \ref{b(r)00}\right)  $, one obtains
\begin{equation}
I_{V}=-\frac{1}{\kappa}\int_{r_{0}}^{\infty}\frac{\left(  \omega+1\right)
r_{0}^{2}}{\omega r^{2}}dr=-\frac{\left(  \omega+1\right)  r_{0}}{\omega
\kappa}. \label{IVo}%
\end{equation}
Even in this case, $I_{V}$ is finite everywhere. The reason is that the
structure of the shape function and of the redshift function are equal to the
pure Casimir wormhole\cite{CW} Therefore we can conclude that, in proximity of
the throat the ANEC can be arbitrarily small.

\subsection{Properties of the Casimir wormhole with an additional
electromagnetic field for the special case $r_{1}^{2}=r_{2}^{2}$}

\label{p3a}In section \ref{p2a}, we have considered the special case in which
the negative Casimir energy is compensated by the positive electromagnetic
field with the assumption that $r_{1}^{2}=r_{2}^{2}$. We want to discuss some
of its properties, even if the size of this TW is Planckian. Repeating the
same steps of Section \ref{p3}, we find that the proper radial distance is%

\begin{gather}
l\left(  r\right)  =\pm\int_{r_{0}}^{r}\frac{dr^{\prime}}{\sqrt{1-\frac{r_{0}%
}{r^{\prime}}}}=\pm\left(  \sqrt{r-r_{0}}\sqrt{r}\right. \nonumber\\
\left.  +\frac{r_{0}}{2}\ln\!\left(  1+2\frac{\sqrt{r-r_{0}}\,\left(  \sqrt
{r}+\sqrt{r-r_{0}}\right)  \,}{\!r_{0}}\right)  \right)  \label{l(r)e}%
\end{gather}
and its asymptotic behavior becomes%
\begin{equation}
l\left(  r\right)  \simeq%
\begin{array}
[c]{ll}%
\pm\left(  r+\frac{r_{0}}{2}\left(  \ln\!\left(  \frac{4r}{r_{0}}\right)
\!-1\right)  +O\left(  \frac{1}{r^{2}}\right)  \right)  & r\rightarrow\infty\\
\pm2\sqrt{r_{0}\left(  r-r_{0}\right)  }+O\left(  \left(  r-r_{0}\right)
^{\frac{3}{2}}\right)  & r\rightarrow r_{0}%
\end{array}
,
\end{equation}
where the\textquotedblleft$\pm$\textquotedblright\ depends on the wormhole
side we are. From Eq.$\left(  \ref{vt}\right)  $ and Eq.$\left(
\ref{vtau}\right)  $, we can compute the total time $\Delta t$ and the proper
total time $\Delta\tau$, respectively. The total time is%
\begin{equation}
\Delta t=\int_{r_{0}}^{r}\frac{\sqrt{r^{\prime}}\exp\left(  -\frac{r_{0}%
}{2r^{\prime}}\right)  }{v\sqrt{r^{\prime}-r_{0}}}dr^{\prime}%
\end{equation}
and it is bounded by the following inequality chain%
\begin{equation}
\frac{1}{v\sqrt{e}}\int_{r_{0}}^{r}\frac{\sqrt{r^{\prime}}}{\sqrt{r^{\prime
}-r_{0}}}dr^{\prime}\leq\Delta t\leq\frac{1}{v}\int_{r_{0}}^{r}\frac
{\sqrt{r^{\prime}}}{\sqrt{r^{\prime}-r_{0}}}dr^{\prime}.
\end{equation}
On the other hand the proper total time is simply%
\begin{equation}
\Delta\tau=\frac{\Delta l}{v}.
\end{equation}
On the same ground, we can compute the embedded surface, which is defined by%
\begin{equation}
z\left(  r\right)  =\pm\int_{r_{0}}^{r}\frac{dr^{\prime}}{\sqrt{\frac
{r^{\prime}}{b\left(  r^{\prime}\right)  }-1}}=\pm2\sqrt{r_{0}}\sqrt{r-r_{0}}%
\end{equation}
Note that in this special case, the total gravitational energy for a
wormhole\cite{NZCP} is vanishing. As regards the acceleration felt by the
traveller, the relationship $\left(  \ref{acc}\right)  $ becomes%
\begin{equation}
\left\vert \mathbf{a}\right\vert =\left\vert \sqrt{1-\frac{r_{0}}{r}}%
\frac{r_{0}}{2r^{2}}\right\vert \leq\frac{g_{\oplus}}{c^{2}},
\end{equation}
where we have assumed a constant speed and $\gamma\simeq1$. Even in this case,
in proximity of the throat, the traveller has a vanishing acceleration. On the
other hand, the radial tidal constraint $\left(  \ref{RTC}\right)  $ and the
lateral tidal constraint $\left(  \ref{LTC}\right)  $ become respectively on
the throat%
\begin{equation}
\frac{r_{0}^{2}}{4r^{4}}\underset{r\rightarrow r_{0}}{=}\frac{1}{4r_{0}^{2}%
}\leq\frac{g_{\oplus}}{c^{2}\left\vert \eta^{\hat{1}^{\prime}}\right\vert
}\qquad\frac{10^{8}m}{2g_{\oplus}}\leq r_{0}\Longrightarrow\qquad
10^{8}m\lesssim r_{0} \label{RTCte}%
\end{equation}
and%
\begin{equation}
\frac{v^{2}}{2r^{2}}\left\vert -\frac{r_{0}}{r}\right\vert \left\vert
\eta^{\hat{2}^{\prime}}\right\vert \lesssim g_{\oplus}\qquad\Longrightarrow
\qquad v\lesssim r_{0}\sqrt{2g_{\oplus}}\qquad\Longrightarrow\qquad
v\lesssim4.43r_{0}\text{ }m/s,
\end{equation}
where we have assumed that the size of the traveller, described by $\eta
^{\hat{1}^{\prime}}$ and $\eta^{\hat{2}^{\prime}}$, is fixed at the symbolic
value of $2$ $m$. If the observer has a vanishing $v$, then the tidal forces
are null. We can use these last estimates to complete the evaluation of the
crossing time which approximately is%
\begin{equation}
\Delta t\simeq\frac{\Delta l}{4.43r_{0}}\simeq4.5\times10^{3}s,
\end{equation}
where we have assumed that $\Delta l\simeq2\times10^{4}r_{0}$ like in Section
\ref{p3}. It is important to observe that since $r_{0}$ has a Planckian value,
then one finds $\left\vert \eta^{\hat{1}^{\prime}}\right\vert \lesssim
2.1\times10^{-43}%
\operatorname{m}%
$. This means that with a Planckian wormhole nothing can traverse it. The last
property we are going to discuss is the \textquotedblleft total
amount\textquotedblright\ of ANEC violating matter in the spacetime\cite{VKD}
which is described by Eq.$\left(  \ref{IV}\right)  $ and for the present case,
one finds%
\begin{equation}
I_{V}=\frac{1}{\kappa}\int\left(  r-r_{0}\right)  \left[  \ln\left(
\frac{e^{-r0/r}}{1-\frac{r_{0}}{r}}\right)  \right]  ^{\prime}dr=-\frac
{1}{\kappa}\int_{r_{0}}^{\infty}\frac{r_{0}^{2}}{r^{2}}dr=-\frac{r_{0}}%
{\kappa},
\end{equation}
Even in this case, $I_{V}$ is finite everywhere and this corresponds to taking
the limit $\omega\rightarrow\infty$ in Eq.$\left(  \ref{IVo}\right)  $.

\section{The Casimir Traversable Wormhole with an Additional Electric Charge:
the constant Plates separation case}

\label{p4}In this section we consider the following setting for the energy
density and radial pressure%
\begin{equation}
\rho_{C}\left(  d\right)  =-\frac{\hbar c\pi^{2}}{720d^{4}};\qquad
p_{r,C}\left(  d\right)  =-\frac{3\hbar c\pi^{2}}{720d^{4}};\qquad\rho
_{E}\left(  r\right)  =\frac{Q^{2}}{2\left(  4\pi\right)  ^{2}\varepsilon
_{0}r^{4}};\qquad p_{r,E}\left(  r\right)  =-\frac{Q^{2}}{2\left(
4\pi\right)  ^{2}\varepsilon_{0}r^{4}}, \label{rhoprd}%
\end{equation}
namely that our exotic matter will be represented by the Casimir energy
density $\left(  \ref{rhoC}\right)  $ and only the electric field is variable
with a radial coordinate $r$. Of course, even in this case we find that the
NEC is violated, namely%
\begin{equation}
\rho\left(  r\right)  +p_{r}\left(  r\right)  =\rho_{C}\left(  d\right)
+p_{r,C}\left(  d\right)  +\rho_{E}\left(  r\right)  +p_{r,E}\left(  r\right)
=-\frac{4\hbar c\pi^{2}}{720d^{4}}<0.
\end{equation}
The total energy density is represented by%
\begin{equation}
\rho\left(  r\right)  =\rho_{C}\left(  d\right)  +\rho_{E}\left(  r\right)
=-\frac{\hbar c\pi^{2}}{720d^{4}}+\frac{Q^{2}}{2\left(  4\pi\right)
^{2}\varepsilon_{0}r^{4}}. \label{rho(r,d)}%
\end{equation}
It is interesting to observe that, differently from Eq. $\left(
\ref{rho(r)}\right)  $, the energy density $\left(  \ref{rho(r,d)}\right)  $
vanishes when%
\begin{equation}
r=\bar{r}=d\sqrt{\frac{r_{2}}{r_{1}}}=\frac{d\sqrt{n}}{\pi}\sqrt[4]%
{90\pi\alpha}, \label{rbar}%
\end{equation}
where $Q=ne$, $e$ is the electron charge, $\alpha$ is the fine structure
constant and $n$ is the total number of the electron charges. In particular,
we find that%
\begin{equation}
\rho\left(  r\right)  \gtreqless0,\qquad when\qquad\left\{
\begin{array}
[c]{c}%
r_{0}\leq r<\bar{r}\\
r=\bar{r}\\
r>\bar{r}%
\end{array}
\right.  .
\end{equation}
The shape function $b\left(  r\right)  $ can be obtained plugging $\rho\left(
r\right)  $ $\left(  \ref{rho(r)}\right)  $ into Eq.$\left(  \ref{rhoEFE}%
\right)  $, whose solution leads to%
\begin{align}
b\left(  r\right)   &  =r_{0}+\frac{GQ^{2}}{2\pi c^{4}\varepsilon_{0}}%
\int_{r_{0}}^{r}\frac{dr^{\prime}}{r^{\prime2}}-\frac{\pi^{3}}{90d^{4}}\left(
\frac{\hbar G}{c^{3}}\right)  \int_{r_{0}}^{r}r^{\prime2}dr^{\prime
}\nonumber\\
&  =r_{0}+\frac{r_{2}^{2}}{r_{0}}-\frac{r_{2}^{2}}{r}-\frac{r_{1}^{2}}{3d^{4}%
}\left(  r^{3}-r_{0}^{3}\right)  , \label{b(r)1}%
\end{align}
where $r_{1}$ and $r_{2}$ have the same meaning of the previous section. We
know that the shape function $\left(  \ref{b(r)1}\right)  $ does not represent
a TW because it is not asymptotically flat. Moreover, for large $r$, $b\left(
r\right)  $ becomes negative. This means that there exists $\tilde{r}$ such
that $b\left(  \tilde{r}\right)  =0$. However, instead of discarding $b\left(
r\right)  $ of Eq. $\left(  \ref{b(r)1}\right)  $, we can try to establish if
there is a way to obtain a TW from Eq.$\left(  \ref{b(r)1}\right)  $. One
important property is the flare out condition described by%
\begin{equation}
b^{\prime}\left(  r_{0}\right)  <1\qquad\Longleftrightarrow\qquad\frac
{r_{2}^{2}d^{4}-r_{0}^{4}r_{1}^{2}}{r_{0}^{2}d^{4}}<1,
\end{equation}
which is satisfied when%
\begin{equation}
r_{0}>\frac{\sqrt{2}d}{2r_{1}}\sqrt{-d^{2}+\sqrt{d^{4}+4r_{1}^{2}r_{2}^{2}}}.
\end{equation}
Another property that has to be satisfied is the absence of a horizon for the
redshift function. From Eq.$\left(  \ref{pr0}\right)  $ and Eq.$\left(
\ref{b(r)1}\right)  $ one finds%
\begin{equation}
\Phi^{\prime}\!\left(  r\right)  =\frac{\left(  3d^{4}r_{0}^{2}+3d^{4}%
r_{2}^{2}+r_{0}^{4}r_{1}^{2}\right)  r-6r_{2}^{2}r_{0}\,d^{4}-10r^{4}r_{1}%
^{2}r_{0}}{2r_{0}r_{1}^{2}r^{5}+6d^{4}r_{0}r^{3}+\left(  -6d^{4}r_{0}%
^{2}-6d^{4}r_{2}^{2}-2r_{0}^{4}r_{1}^{2}\right)  r^{2}+6d^{4}r_{0}r_{2}^{2}r}.
\label{Phi'}%
\end{equation}
Close to the throat, the r.h.s. can be approximated by%
\begin{equation}
\Phi^{\prime}\!\left(  r\right)  =\frac{\left(  3d^{4}r_{0}^{2}+3d^{4}%
r_{2}^{2}+r_{0}^{4}r_{1}^{2}\right)  r_{0}-6r_{2}^{2}r_{0}d^{4}-10r_{0}%
^{5}r_{1}^{2}}{10r_{0}^{5}r_{1}^{2}+18d^{4}r_{0}^{3}+2\left(  -6d^{4}r_{0}%
^{2}-6d^{4}r_{2}^{2}-2r_{0}^{4}r_{1}^{2}\right)  r_{0}+6r_{2}^{2}r_{0}d^{4}%
}\left(  r-r_{0}\right)  ^{-1}+\mathrm{O}\!\left(  1\right)  .
\end{equation}
It is immediate to see that a horizon will be present, unless we impose that%
\begin{equation}
\left(  3d^{4}r_{0}^{2}+3d^{4}r_{2}^{2}+r_{0}^{4}r_{1}^{2}\right)
r_{0}-6r_{2}^{2}r_{0}d^{4}-10r_{0}^{5}r_{1}^{2}=0.
\end{equation}
We have four solutions, but only two are real. They are represented by%
\begin{equation}
r_{0}=\frac{\sqrt{6}\,\sqrt{d^{2}\pm\sqrt{d^{4}-12r_{1}^{2}r_{2}^{2}}}%
\,d}{6r_{1}}. \label{throat}%
\end{equation}
Among them, the first one is%
\begin{equation}
r_{0}=\frac{\sqrt{6}d}{6r_{1}}\sqrt{d^{2}+\sqrt{d^{4}-12r_{1}^{2}r_{2}^{2}}%
}\underset{d\gg r_{1}}{\simeq}\frac{\sqrt{3}d^{2}}{3r_{1}}+O\left(  \frac
{1}{d^{2}}\right)  , \label{r01}%
\end{equation}
which is independent on $r_{2}$ and therefore on the electric field. The
result has a dependence on $d$ similar to the one obtained in
Refs.\cite{GABTW,YCW}. The other interesting solution is%
\begin{equation}
r_{0}=\frac{\sqrt{6}d}{6r_{1}}\sqrt{d^{2}-\sqrt{d^{4}-12r_{1}^{2}r_{2}^{2}}%
}\underset{d\gg r_{1}}{\simeq}r_{2}+O\left(  \frac{1}{d^{4}}\right)  ,
\label{r02}%
\end{equation}
which is independent on the plates separation but it is depending on the
electric field. However, the two solutions can be merged into one if%
\begin{equation}
d=\sqrt[4]{3}\sqrt{2r_{1}r_{2}}. \label{d}%
\end{equation}
It is straightforward to verify that%
\begin{equation}
\frac{\sqrt{6}\,\sqrt{d^{2}\pm\sqrt{d^{4}-12r_{1}^{2}r_{2}^{2}}}\,d}{6r_{1}%
}>\frac{\sqrt{2}d}{2r_{1}}\sqrt{-d^{2}+\sqrt{d^{4}+4r_{1}^{2}r_{2}^{2}}}%
\end{equation}
even when the condition $\left(  \ref{d}\right)  $ is used. This means that
the throat location is compatible with the flare out condition. Note that
Eq.$\left(  \ref{d}\right)  $ establishes a relationship between the plates
separation and the charge quantity that can be used. In particular, if $d$ is
of the order of $nm$ $\left(  d\simeq10^{-9}m\right)  $ and $r_{1}$ is
Planckian, one gets%
\begin{equation}
\frac{d}{r_{1}}\simeq\frac{10^{-9}m}{10^{-35}m}\simeq\frac{n}{\pi}\sqrt
{\frac{180}{137\pi}}\simeq\allowbreak0.205\,85n\qquad\Longrightarrow\qquad
n\simeq10^{26}, \label{nrs}%
\end{equation}
where we have used the fine structure constant introduced in $\left(
\ref{fine}\right)  $. Note that, for silver the average number of conduction
electrons is $5.8\cdot10^{28}/m^{3}$. As we can see, the relationship $\left(
\ref{d}\right)  $ constrains the plates separation to be a function of the
electric charge. This also implies that the throat $\left(  \ref{throat}%
\right)  $ reduces to%
\begin{equation}
r_{0}=\sqrt{2}r_{2}. \label{r0d}%
\end{equation}
If we put numbers in $\left(  \ref{r0d}\right)  $, one finds%
\begin{equation}
r_{0}=\frac{180n^{2}}{137\pi^{3}}r_{1}\simeq\left(  4.\,\allowbreak
237\,4\times10^{-2}n^{2}\right)  \left(  1.6\times10^{-37}m\right)
\simeq\left(  10^{26}\right)  ^{2}6.\,\allowbreak779\,8\times10^{-39}%
m\simeq10^{13}m. \label{r0n}%
\end{equation}
Note that the equation connecting the throat with the plates separation and
the fine structure constant can be obtained also with the help of an EoS of
the form $p_{r}\left(  r\right)  =\omega\left(  r\right)  \rho\left(
r\right)  $ with%
\begin{equation}
\omega\left(  r\right)  =-\frac{b\left(  r\right)  }{rb^{\prime}\left(
r\right)  }=rd^{4}\left(  \frac{r_{0}+\frac{r_{2}^{2}}{r_{0}}-\frac{r_{2}^{2}%
}{r}-\frac{r_{1}^{2}}{3d^{4}}\left(  r^{3}-r_{0}^{3}\right)  }{r_{1}^{2}%
r^{4}-r_{2}^{2}d^{4}}\right)  \label{omega(r)1}%
\end{equation}
allowing to fix $\Phi^{\prime}\!\left(  r\right)  =0$. On the other hand, from
$p_{r}\left(  r\right)  $ and $\rho\left(  r\right)  $ defined in $\left(
\ref{rhoprd}\right)  $, one can obtain also%
\begin{equation}
\omega\left(  r\right)  =\frac{p_{r}\left(  r\right)  }{\rho\left(  r\right)
}=\frac{3r_{1}^{2}r^{4}+r_{2}^{2}d^{4}}{r_{1}^{2}r^{4}-r_{2}^{2}d^{4}}.
\label{omega(r)2}%
\end{equation}
If we impose that $\omega\left(  r\right)  $ in Eq.$\left(  \ref{omega(r)1}%
\right)  $ be equal to $\omega\left(  r\right)  $ in Eq.$\left(
\ref{omega(r)2}\right)  $, one finds that the only solution is represented by
Eq.$\left(  \ref{throat}\right)  $. Nevertheless, outside of the throat, the
function $\omega\left(  r\right)  $ in Eq.$\left(  \ref{omega(r)1}\right)  $
is no longer equal to the one of Eq.$\left(  \ref{omega(r)2}\right)  $.
Plugging the value of $r_{0}$ in $\left(  \ref{r0d}\right)  $ into the shape
function $\left(  \ref{b(r)1}\right)  $, one finds%
\begin{equation}
b\left(  r\right)  =\frac{14}{9}r_{0}-\frac{r_{0}^{2}}{2r}-\frac{r^{3}%
}{18r_{0}^{2}}. \label{b(r)1r}%
\end{equation}
In this reduced form, it is easier to see that there exists%
\begin{equation}
\tilde{r}=2.9208r_{0}\qquad where\qquad b\left(  \tilde{r}\right)  =0.
\label{zero}%
\end{equation}
For $r>\bar{r}$, $b\left(  r\right)  <0$. Therefore to avoid negative values,
we can cut off the region where $r>\bar{r}$. For this reason, the range of the
wormhole must be constrained to be very close to the throat. A possible
profile comes from the following setup%
\begin{equation}
\left\{
\begin{array}
[c]{l}%
b\left(  r\right)  =\frac{14}{9}r_{0}-\frac{r_{0}^{2}}{2r}-\frac{r^{3}%
}{18r_{0}^{2}}\qquad r_{0}\leq r\leq\tilde{r}\qquad\qquad b\left(  r\right)
=0\qquad r\geq\tilde{r}\\
\Phi\!\left(  r\right)  =0.
\end{array}
\right.  \label{ABTW1}%
\end{equation}
Note that in the region $\bar{r}\leq r\leq\tilde{r}$, the constant Casimir
source becomes relevant. One can be tempted to classify such a TW as an
Absurdly Benign Traversable Wormhole (ABTW) defined by\cite{GABTW}%
\begin{equation}
\left\{
\begin{array}
[c]{l}%
b\left(  r\right)  =r_{0}\left(  1-\mu\left(  r-r_{0}\right)  \right)
^{2}\qquad r_{0}\leq r\leq r_{0}+\frac{1}{\mu};\qquad\qquad b\left(  r\right)
=0\qquad r\geq r_{0}+\frac{1}{\mu}\\
\Phi\!\left(  r\right)  =0.
\end{array}
\right.
\end{equation}
Even if, outside the region $r=\bar{r}$, $\rho\left(  r\right)  $ and
$p_{r}\left(  r\right)  $ tend to a constant value, one has to realize that
this behavior cannot be extended to the whole space, rather it is likely that
$\rho\left(  r\right)  $ and $p_{r}\left(  r\right)  $ vanish in proximity of
the external part of the plates, like in the model introduced by Visser who
proposed to consider the following SET\cite{Visser}%
\begin{gather}
T_{\sigma}^{\mu\nu}=\sigma\hat{t}^{\mu}\hat{t}^{\nu}\left[  \delta\left(
z\right)  +\delta\left(  z-a\right)  \right] \nonumber\\
+\Theta\left(  z\right)  \Theta\left(  a-z\right)  \frac{\hbar c\pi^{2}%
}{720a^{4}}\left[  \eta^{\mu\nu}-4\hat{z}^{\mu}\hat{z}^{\nu}\right]  ,
\label{TCas}%
\end{gather}
where $\hat{t}^{\mu}$ is a unit time-like vector, $\hat{z}^{\mu}$ is a normal
vector to the plates and $\sigma$ is the mass density of the plates.

\section{Conclusions}

\label{p5}In this paper, we have extended the study begun in Ref.\cite{CW} by
including an electromagnetic source. Since the electromagnetic field satisfies
the property $\left(  \ref{prho}\right)  $, the NEC is still violated and it
seems to be independent on the strength of the electromagnetic field.
Repeating the same strategy adopted in Ref.\cite{CW}, we have found another
Casimir wormhole with a different $\omega$ as it should be. We would like to
draw the reader's attention that the additional electric field is a part of
the source and not a feature of the TW. The most important consequence is that
the wormhole throat becomes directly dependent on the charge in an additive
way, even if under the square root%
\begin{equation}
r_{0}=\sqrt{3r_{1}^{2}+r_{2}^{2}}.
\end{equation}
If this result seems to be encouraging, on the other hand we have two aspects
that must be explored. The first one is that for $r_{2}\gg r_{1}$, the energy
density becomes positive. At this stage of the analysis, we do not know if the
TW ceases to exist or not. A possible answer could come from a back reaction
investigation of the electromagnetic and gravitational fields together which
is beyond the purpose of this paper. The second aspect is that, always in the
range where $r_{2}\gg r_{1}$, $\omega\rightarrow-1$ and a horizon seems to be
appear. However, this is the result of a limiting procedure and the value
$\omega=-1$ can never be reached. To this purpose, we have to recall that, it
is the NEC that must be violated. This means that with the help of
\textquotedblleft\textit{phantom energy}\textquotedblright, $\rho\left(
r\right)  >0$\cite{phantomWH1,phantomWH2,phantomWH3}. Indeed, the following
relationship%
\begin{gather}
p_{r}\left(  r\right)  =\omega\rho\left(  r\right)  ,\qquad\Longrightarrow
\qquad p_{r}\left(  r\right)  +\rho\left(  r\right)  <0\nonumber\\
\qquad\Longleftrightarrow\qquad\left(  1+\omega\right)  \rho\left(  r\right)
<0,
\end{gather}
allows to keep $\rho\left(  r\right)  >0$, provided that $\omega<-1$. However,
generally speaking, it is not known how to build and manipulate such a
\textquotedblleft\textit{phantom energy}\textquotedblright. The electrovacuum
example we have discussed in this paper seems to be encouraging, because in
this context, the electromagnetic field seems to behave exactly like a phantom
source. Note that this is not true when the electromagnetic field is the only
source. To further proceed, note that there is an essential discontinuity when
$r_{2}=r_{1}$ into the relationship $\left(  \ref{or1r2}\right)  $. For this
reason this case has been examined separately in paragraph \ref{p2a}. It is
important to observe that even in this case, a TW can exist at zero density
with a throat of Planckian size. A different behavior appears when one
considers the mixed source case, namely only the electromagnetic field has a
variable radius, while the plates separation has been dealt as a parameter. In
this framework, one finds that it is possible to avoid the creation of a
horizon if the throat satisfies Eq.$\left(  \ref{throat}\right)  $. Because we
have two solutions, it is possible to have only one solution, if we impose the
constraint $\left(  \ref{d}\right)  $. This constraint creates a relationship
between the plates separation, the Planck length which is not modifiable and
the quantity of charge introduced which can be changed. Unfortunately, even in
this case the main limitation comes from the plates separation which has
consequences also on the throat size. However, such a limitation is not so
stringent like in Refs.\cite{GABTW,YCW} because of the presence of the square
root in Eq.$\left(  \ref{d}\right)  $. Indeed, if it could be possible to push
the plates separation at a distance of the order of $pm$, one could find that
the throat could be of the order of $10^{9}m$: a gain of a factor $10^{2}$
with respect to what we have found in Ref.\cite{GABTW}. Note that the
constraint $\left(  \ref{d}\right)  $ implies that we have a throat radius
directly proportional to the charge of the electromagnetic field. For this
reason, differently from the pure Casimir source, the introduction of an
electromagnetic field seems to go towards a traversable wormhole which is a
little less traversable in principle and a little more traversable in
practice. It is important also to observe that the shape function $\left(
\ref{b(r)1}\right)  $ and subsequently the shape function $\left(
\ref{b(r)1r}\right)  $ can be promoted to be a traversable wormhole shape
function if we assume that there is a smooth transition between the curved
space and flat space expressed by Eq.$\left(  \ref{zero}\right)  $.
Alternatively, one could use the \textit{cut and paste} technique and glue the
shape function $\left(  \ref{b(r)1r}\right)  $ with a Schwarzschild profile.
Of course the whole analysis can be generalized to include even a magnetic
field and this will be examined in a future publication. It is interesting to
observe that contrary to the pure Casimir wormhole, the additional
electromagnetic field avoids that the TW has a throat of Planckian size.
Needless to say that things could drastically change with the inclusion of
quantum corrections on the line of Refs.\cite{RG,RG1,RGFSNL,RGFSNL1,RGFSNL2}.

\end{document}